\documentclass[prl,twocolumn,preprintnumbers,floatfix,amsmath,amssymb]{revtex4}

\usepackage{graphicx}
\usepackage{latexsym}
\usepackage{amsmath}
\usepackage{amssymb}
\usepackage{amsfonts}
\usepackage{bm}

\newcommand{\rvec}{\ensuremath{\underline{r}}}

\newcommand{\Rend}{\ensuremath{R_\mathrm{e}}}
\newcommand{\Rgyr}{\ensuremath{R_\mathrm{g}}}
\newcommand{\rhostar}{\ensuremath{\rho^*}}
\newcommand{\dperi}{\ensuremath{d_\mathrm{p}}}

\newcommand{\ninter}{\ensuremath{n_\mathrm{int}}}
\newcommand{\ginter}{\ensuremath{g_\mathrm{int}}}

\newcommand{\thetatwo}{\ensuremath{\theta_\mathrm{2}}}

\newcommand{\Rdil}{\ensuremath{R_\mathrm{0}}}
\newcommand{\nudil}{\ensuremath{\nu_\mathrm{0}}}

\begin{document}
\title{Interchain monomer contact probability in two-dimensional polymer solutions} 

\author{N.~Schulmann}
\affiliation{Institut Charles Sadron, Universit\'e de Strasbourg, CNRS, 23 rue du Loess, 67034 Strasbourg Cedex, France}
\author{H. Meyer}
\affiliation{Institut Charles Sadron, Universit\'e de Strasbourg, CNRS, 23 rue du Loess, 67034 Strasbourg Cedex, France}
\author{J.P.~Wittmer}
\email{joachim.wittmer@ics-cnrs.unistra.fr}
\affiliation{Institut Charles Sadron, Universit\'e de Strasbourg, CNRS, 23 rue du Loess, 67034 Strasbourg Cedex, France}
\author{A. Johner}
\affiliation{Institut Charles Sadron, Universit\'e de Strasbourg, CNRS, 23 rue du Loess, 67034 Strasbourg Cedex, France}
\author{J. Baschnagel}
\affiliation{Institut Charles Sadron, Universit\'e de Strasbourg, CNRS, 23 rue du Loess, 67034 Strasbourg Cedex, France}

\begin{abstract} 
Using molecular dynamics simulation of a standard bead-spring model we investigate the density crossover 
scaling of strictly two-dimensional ($d=2$) self-avoiding polymer chains focusing on properties related 
to the contact exponent $\thetatwo$ set by the intrachain subchain size distribution. With $R \sim N^{\nu}$ 
being the size of chains of length $N$, the number $\ninter$ of interchain monomer contacts per monomer 
is found to scale as $\ninter \sim 1/N^{\nu \thetatwo}$ with $\nu=3/4$ and $\thetatwo=19/12$ for 
dilute solutions and $\nu=1/d$ and $\thetatwo=3/4$ for $N \gg g(\rho) \approx 1/\rho^2$.
Irrespective of the density $\rho$ sufficiently long chains are thus found to consist of 
compact packings of blobs of fractal perimeter dimension $\dperi = d-\thetatwo = 5/4$.
\end{abstract}

\date{\today}
\maketitle

As first suggested by de Gennes \cite{deGennesBook}, it is now generally accepted
\cite{Duplantier89,ANS03,mai99,mai00,Deutsch05,Rubinstein07,carmesin90,Rutledge97,Yethiraj03,MKA09,MWK10,MSZ11}
that strictly two-dimensional (2D) self-avoiding polymer chains 
adopt compact and segregated conformations at high densities, i.e.,
the typical chain size $R$ scales as
\begin{equation}
R(N,\rho) \approx (N/\rho)^{\nu} \mbox{ where } \nu=1/d=1/2
\label{eq_compact}
\end{equation}
with $N$ being the chain length, $\rho$ the monomer number density,
and $d=2$ the spatial dimension. We assume here that monomer overlap and chain
intersection are strictly forbidden \cite{ANS03}.
Compactness obviously does {\em not} imply Gaussian chain statistics nor does segregation of chains
impose disklike shapes minimizing the contour perimeter of the chains. Focusing on dense 2D melts 
it has been shown in fact \cite{ANS03,MKA09,MWK10,MSZ11} that the irregular chain contours are 
characterized by a fractal perimeter of typical length 
\begin{equation}
L \sim N \ninter \sim R^{\dperi} \sim N^{\dperi/d}
\mbox{ with } \dperi = d - \thetatwo = 5/4
\label{eq_dperi}
\end{equation}
where $\ninter$ stands for the fraction of monomers of a chain interacting with monomers of 
other chains. The fractal line dimension $\dperi$ is set by Duplantier's contact exponent
$\thetatwo=3/4$ characterizing the size distribution of inner chain segments \cite{Duplantier89}. 
Since the melt limit is experimentally difficult to realize for a strictly 2D layer 
\cite{mai99,mai00,Deutsch05}, the aim of this {\em Letter} is to show that eq~\ref{eq_dperi}
holds more generally for all densities if the chains are sufficiently long.

\begin{figure}[t]
\centerline{\resizebox{0.95\columnwidth}{!}{\includegraphics*{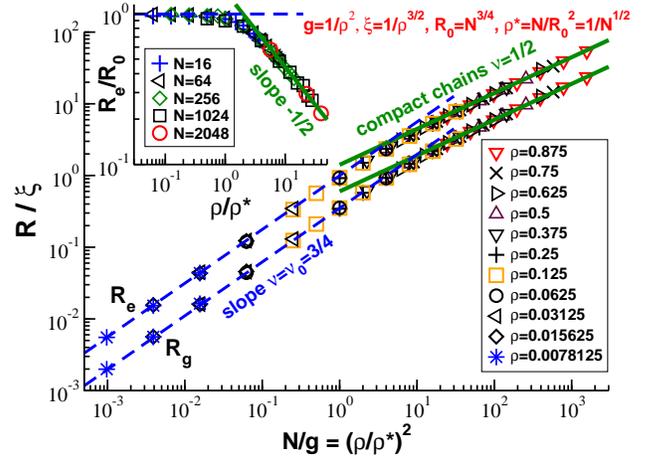}}}
\caption{Average chain size $R$ for self-avoiding polymers in $d=2$ dimensions
for a broad range of densities $\rho$ and chain lengths $N$. 
The exponent $\nu=\nudil \equiv 3/4$ expected for the dilute swollen chain limit is given by 
the dashed lines, the compact chain exponent $\nu=1/d$ by the bold lines.
Main panel: Root-mean square end-to-end distance $\Rend/\xi$
and radius of gyration $\Rgyr/\xi$ as a function of reduced chain length $N/g$
where we set $\xi \equiv 1/\rho^{3/2}$ and $g \equiv 1/\rho^2$.
Inset: Replot focusing on the $\rho$-scaling for constant chain length $N$.
\label{fig1}
}
\end{figure}

As described in Ref.~\cite{MWK10} our numerical results are obtained by molecular dynamics simulations 
of monodisperse, linear and highly flexible chains using the well-known Kremer-Grest bead-spring model 
\cite{KG90}. Lennard-Jones (LJ) units are used throughout this paper and temperature is set to unity. 
Note that monomer overlap and chain crossings are made impossible by the excluded volume interaction 
and the bonding potential; the chains are thus strictly 2D ``self-avoiding walks" in the sense of
Ref.~\cite{ANS03}.
While our previous studies \cite{MKA09,MWK10,MSZ11} have focused on one melt density, $\rho = 0.875$, 
we scan now over a broad range of densities $\rho$ as may be seen from Figure~\ref{fig1}. 
Chains of length $N=16$ up to $N=1024$ monomers have been sampled for all $\rho$;
for $\rho=0.125$, $\rho=0.5$ and $\rho=0.875$ we have even computed systems with $N=2048$  
\cite{foot_simu}.

That sufficiently long 2D polymer chains become indeed compact for all densities, 
as stated by eq~\ref{eq_compact}, is shown in Figure~\ref{fig1} presenting the 
overall chain size $R$ as characterized by the root-mean square end-to-end distance 
$\Rend$ and the radius of gyration $\Rgyr$ \cite{DoiEdwardsBook}. 
As one expects \cite{deGennesBook,Duplantier89}, the typical chain size is found to increase with a power-law 
exponent $\nu=\nudil \equiv 3/4$ in the dilute limit (dashed lines) and with $\nu=1/d$ for larger
chains and densities in agreement with various numerical \cite{carmesin90,Rutledge97,Yethiraj03}
and experimental studies \cite{mai99,mai00,Rubinstein07}. 
The main panel presents data for different densities as a function
of chain length $N$ which are successfully brought to collapse by rescaling the horizontal
axis with the number of monomers $g(\rho) \approx \rho \xi^d \sim 1/\rho^{1/(\nudil d -1)}= \rho^{-2}$ 
spanning the semidilute blob and the vertical axis by its size $\xi \sim g^{\nudil} \sim 1/\rho^{3/2}$.
For simplicity, all prefactors related to the dilute limit are set to unity,
e.g. $g\equiv 1/\rho^2$ and $\xi=1/\rho^{3/2}$ \cite{foot_blobprefactor}.
An equivalent data representation focusing on the chain size as a function of density taking as reference
the dilute chain size $\Rdil \equiv N^{\nudil}$ is shown in the inset. We trace here $\Rend/\Rdil$
as a function of the reduced density $\rho/\rhostar$ with $\rhostar \equiv N/\Rdil^d = 1/\sqrt{N}$
being the crossover density. The typical chain size decreases indeed as $R \sim 1/\sqrt{\rho}$ with 
increasing density (bold line) as expected for compact chains, eq~\ref{eq_compact}.
Below we refer to $N/g = (\rho/\rhostar)^2 \gg 1$ as the compact chain limit
and to $g \to 0$ as the melt limit.

\begin{figure}[t]
\centerline{\resizebox{0.95\columnwidth}{!}{\includegraphics*{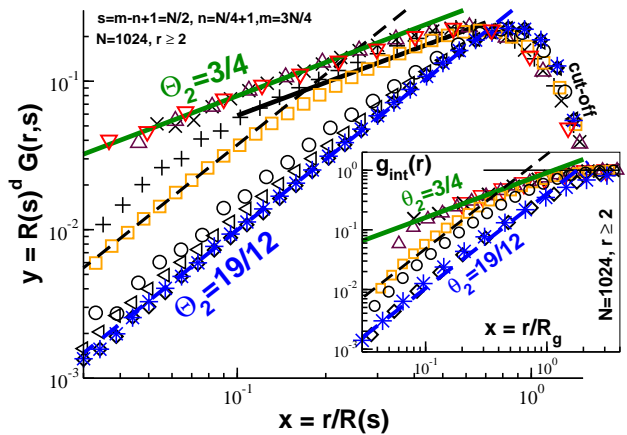}}}
\caption{Main panel:
Subchain size distribution $G(r,s)$ between the monomers $n = N/4+1$ and $m  = n+s-1 = 3N/4$
for $N=1024$ and various densities using the same symbols as in the main panel of Figure~\ref{fig1}.
Motivated by the scaling in the dilute and melt limits, eq~\ref{eq_G2scal}, 
we trace $R^d(s) G(r,s)$ {\em vs.} $x=r/R(s)$ with $R^2(s)$ being the second moment of $G(r,s)$. 
Inset: Interchain monomer pair distribution function $\ginter(r)$ as a function of $x=r/\Rgyr(N)$,
revealing for $x \ll 1$ the same exponents $\thetatwo$ as the intrachain distribution $G(r,s)$.
\label{fig2}
}
\end{figure}

Obviously, the mere fact that $\nu$ becomes $1/2$ in the compact chain limit
does not imply Gaussian chain statistics \cite{Duplantier89,ANS03}. This can be
directly seen, e.g., from the subchain size distribution $G(r,s)$ of the
vector $\rvec = \rvec_m-\rvec_n$ between the monomers $n$ 
and $m=n+s-1$ on the same tagged chain presented in the main panel of Figure~\ref{fig2}
\cite{WCX11}.
For the examples indicated we have chosen subchains of length $s=N/2$ in
the middle of chains of length $N=1024$ \cite{foot_othertheta}.
Only distances $r \ge 2$ are presented to take off local physics related to the LJ beads.
Using the measured second moment $R^2(s)$ of the distribution we trace
$R^d(s) G(r,s)$ as a function of the reduced distance $x = r/R(s)$. 
This is motivated by the scaling
\begin{equation}
R^{d}(s) G(r,s) = f(x)
\mbox{ with } f(x) \approx x^{\thetatwo} \mbox{ for } x \ll 1 
\label{eq_G2scal}
\end{equation}
observed as expected in the dilute and melt limits where $R(s)$ is the only relevant length scale. 
Note that $f(x)$ is a cut-off function for $x \gg 1$ \cite{MWK10}.
The dashed and bold power-law slopes indicated in the figure correspond to the exponents 
$\thetatwo = 19/12$ and $\thetatwo=3/4$ predicted for both limits \cite{Duplantier89}. 
(We remind that for Gaussian chains $\thetatwo \equiv 0$, i.e. $G(r,s)$ decays monotonously 
with distance $r$.)
Obviously, the scaling of $G(r,s)$ becomes more intricate for semi-dilute densities
where the blob size $\xi$ sets an additional length scale.
For $s \gg g(\rho) \gg 1$ one expects to observe two power-law asymptotes,
one scaling as $G(r,s) \sim r^{19/12}$ if the conformational properties within the blob are probed
($r \ll \xi$) and one as $G(r,s) \sim r^{3/4}$ for $\xi \ll r \ll R(s)$. 
Although much larger blobs are certainly warranted to demonstrate this unambiguously,
this expected behavior is qualitatively consistent with our data as may be seen for $\rho=0.125$
\cite{foot_othertheta}.

As shown in the inset of Figure~\ref{fig2}, the $\thetatwo$ exponents do not only
characterize intrachain properties but are also relevant for describing the
interchain monomer contacts. We have plotted here the radial pair distribution
function $\ginter(r)$ between monomers from different chains which becomes
unity (by normalization) for large reduced distances $x = r/\Rgyr \gg 1$.
The scaling of the horizontal axis is again motivated by the density limits 
where the chain size $R$ is the only relevant length scale. 
This scaling can be verified by comparing different chain lengths $N$
at same density $\rho$ (not shown). The power-law exponents $\thetatwo$ for $x\ll 1$ 
can be understood by generalizing a scaling argument given in Ref.~\cite{ANS03}
stating that it is irrelevant of whether the neighborhood at $r \ll R$
around the reference monomer is penetrated by a long loop of the same chain or by
another chain with a center of mass at a typical distance $R$. 
Hence \cite{foot_IntraInter}, 
\begin{equation}
\ginter(r) \rho \approx G(r,s\approx N) \times N \times \underline{R^d\rho/N}
\label{eq_G2ginter}
\end{equation}
where the second factor $N$ stems from the fact that it is irrelevant
which monomer of the second chain is probed.
Note that the underlined term characterizing the probability for having a second chain
close to the reference chain drops out in the compact chain limit, eq~\ref{eq_compact}.
Equations~\ref{eq_G2scal} and \ref{eq_G2ginter} imply
\begin{equation}
\ginter(r) \approx x^{\thetatwo} \sim 1/N^{\nu \thetatwo}
\mbox{ for } x \ll 1
\label{eq_ginterasymp}
\end{equation}
as indicated by dashed and bold solid lines for, respectively, dilute and melt limits.

\begin{figure}[t]
\centerline{\resizebox{0.95\columnwidth}{!}{\includegraphics*{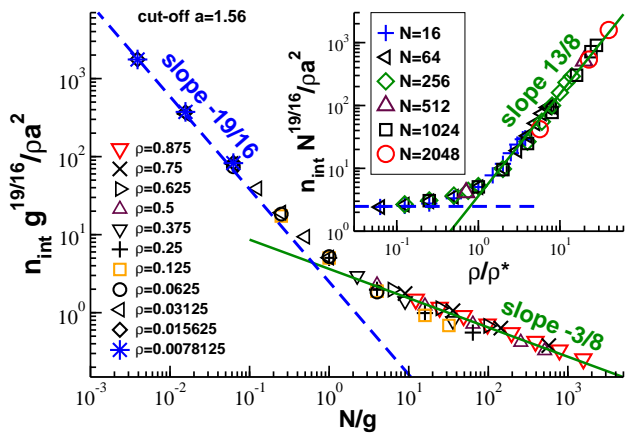}}}
\caption{Average interchain monomer contact number $\ninter$ {\em vs.} 
reduced chain length $N/g$ (main panel) and reduced density $\rho/\rhostar$ (inset). 
The dashed lines correspond to the exponent $\nu \thetatwo = 19/16$ for dilute chains,
the full lines to $\nu \thetatwo = 3/8$ for the compact limit. 
Confirming eqs~\ref{eq_ginterasymp} and \ref{eq_nintercompact}
the observed successful data collapse is the central result of this work.
\label{fig3}
}
\end{figure}

In order to characterize more systematically the density crossover 
we focus now on $\ginter(r \approx a)$ with $a$ being one fixed monomeric distance for all densities. 
For the data presented in Figure~\ref{fig3} we have determined the typical number $\ninter$ of 
interchain monomer contacts for a reference monomer by integrating $\ginter(r) \rho$ 
up to a cut-off $a=1.56$ \cite{foot_acutoff}.
Similar scaling results have been obtained by computing the
fraction of monomers with at least one interchain 
contact within a distance $a$ following Ref.~\cite{MWK10}.
Using a similar representation as in Figure~\ref{fig1}, 
we present the rescaled number of interchain contacts 
$\ninter$ as a function of the reduced chain length $N/g$ for different densities (main panel)
and as a function of the reduced density $\rho/\rhostar$ for different chain lengths (inset). 
The rescaling used for the vertical axes is readily obtained by matching
at $N/g= (\rho/\rhostar)^2 = 1$ the power-law results implied by eq~\ref{eq_ginterasymp}
for the dilute ($\nu \thetatwo = 19/16$) and compact ($\nu \thetatwo = 3/8$) limits.
This implies for the latter limit (solid bold lines) 
\begin{equation}
\ninter \approx \frac{\rho a^d}{g^{19/16}} \times (N/g)^{-3/8}
\label{eq_nintercompact}
\end{equation}
where the first factor corresponds to the fraction of monomers of a given blob interacting 
with monomers from other chains given that the blob is at the chain perimeter.
The second term stands for the fraction of blobs of a chain interacting with other chains
and the successful scaling presented in Figure~\ref{fig3}, thus demonstrates that in the compact
chain limit the perimeter length $L$ scales as $L/\xi \approx (N/g)^{1- \thetatwo/d}$ 
\cite{foot_blobsurface}.
Since on the other hand $L/\xi \approx (R/\xi)^{\dperi} \approx (N/g)^{\dperi/d}$, 
this implies a fractal line dimension $\dperi = d- \thetatwo=5/4$, generalizing thus
the numerical result obtained for the melt limit \cite{MWK10} to the broader compact chain limit.

\begin{figure}[t]
\centerline{\resizebox{0.95\columnwidth}{!}{\includegraphics*{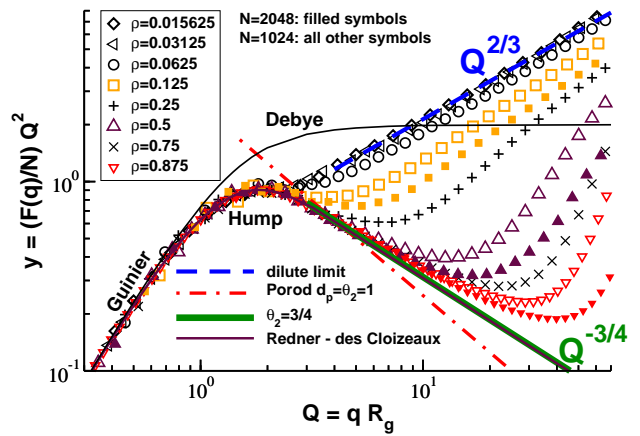}}}
\caption{Kratky representation of the intramolecular form factor $F(q)$.
The Debye formula is indicated by the top thin line,
the dilute swollen chain limit with $F(q) \sim 1/q^{1/\nudil}$ by the dashed line,
the Porod scattering for a compact 2D object with smooth perimeter
by the dash-dotted line and
the power-law exponent $-\thetatwo=-3/4$ by the bold solid line.
Our complete theoretical prediction for asymptotically long chains 
using the so-called Redner-des Cloizeaux approximation of $G(r,s)$ \cite{MWK10}, 
allowing to fit the data starting from the Guinier regime for small wave vectors,
is given by the thin solid line. 
\label{fig4}
}
\end{figure}

Although direct real-space visualizations of 2D polymer systems are experimentally
feasible \cite{mai99,mai00,Rubinstein07}, allowing thus a direct computation of $G(r,s)$ 
and $\ginter(r)$, it is important to note that the exponent $\thetatwo=3/4$ for compact 
chains of blobs may in principle be obtained from an analysis of the intrachain molecular 
form factor $F(q)$ with $q$ being the wave vector \cite{DoiEdwardsBook}. 
This point is addressed in Figure~\ref{fig4} where the rescaled form factor $(F(q)/N) Q^2$
is plotted as a function of the reduced wave vector $Q = q \Rgyr$ for different densities $\rho$.
Filled symbols refer to data for $N=2048$, all other symbols to $N=1024$.
At variance to the indicated Debye formula for Gaussian chains a strong nonmonotonous 
behavior is revealed by our data, which approaches with increasing $\rho$ and $N$ a power-law 
exponent $-\thetatwo=-3/4$ (bold solid line).
This power law limit is implied by the fact that $F(q)$ can be expressed as an integral
over the Fourier transform $G(q,s)$ of the intrachain size distribution $G(r,s)$ 
\cite{MWK10,WCX11}. 
Using general arguments which do not depend explicitly on the density or the blob size 
it has been shown \cite{MWK10} that as a direct consequence of eq~\ref{eq_G2scal} 
the form factor for asymptotically long chains must scale as
\begin{equation}
\frac{F(q)}{N} Q^d \approx \frac{1}{Q^{\thetatwo}} \approx \frac{1}{Q^{d-\dperi}}
\mbox{ for } \frac{1}{R} \ll q \ll \frac{1}{\xi}
\label{eq_Fqpower}
\end{equation}
where in the last step of the equation chain we have used $\dperi=d-\thetatwo$. 
We emphasize that this result is consistent with the well-known Porod scattering expected 
for general compact objects of fractal surface dimension $\dperi$ \cite{MSZ11}. 
Obviously, much larger chains as computationally feasible at present are required to confirm 
numerically the predicted perimeter dimension using eq~\ref{eq_Fqpower}.

In summary, we have investigated the density crossover scaling of strictly 2D self-avoiding 
polymer chains focusing on properties related to the $\thetatwo$ contact exponent which  
characterizes not only (by definition) the intrachain distribution $G(r,s)$,
but also the interchain pair distribution function $\ginter(r) \sim r^{\thetatwo}$
and the monomer contact number $\ninter \sim 1/N^{\nu\thetatwo}$. 
The established scaling for dense melts \cite{ANS03,MKA09,MWK10} is found to
remain valid for all densities if reformulated in terms of chains of blobs.
As a consequence, the intramolecular form factor $F(q)$ measuring the composition fluctuations 
at the fractal perimeters of the compact blob chains approaches with increasing density
and chain length a power-law asymptote implied by the generalized Porod scattering
of compact objects of fractal perimeter dimension $\dperi=d-\thetatwo=5/4$.
Note that our chains are sufficiently long to confirm $\dperi$
using the intermolecular binary contacts $\ninter$ (Figure~\ref{fig3})
but not using the form factor. One reason for this is that eq~\ref{eq_Fqpower}
assumes asymptotically long chains where critical exponents other than $\thetatwo$
may be ignored. The corrections for finite $N/g$ need still to be worked out theoretically.
However, since $\ginter(r)$ or $\ninter$ may be directly obtained experimentally by
visualization of dilute and semidilute 2D polymer solutions \cite{mai99,mai00,Rubinstein07},
our numerical results suggest that the analysis of the latter properties may allow 
the experimental verification of the predicted exponents. 

Finally, we emphasize that at variance to most experimental systems \cite{mai99,mai00,Rubinstein07} 
we have assumed in the presented work that the chains are flexible down to monomeric scales. 
As long as the persistence length remains smaller than the blob size, this should 
not alter the suggested scaling properties. In fact, since the blob size decreases very 
strongly with persistence length, i.e. $N/g$ increases, a finite rigidity might even speed 
up the convergence to the predicted asymptotic behavior.
In order to bring our computational approach closer to experiment
we are currently computing systems with finite persistence length.

\begin{acknowledgments}
We thank for generous grants of computer time by GENCI-IDRIS (Grant No. i2009091467)
and the P\^ole Mat\'eriaux et Nanosciences d'Alsace.
N.S. acknowledges financial support by the R\'egion Alsace.
We are indebted to T. Kreer (Dresden) and M. Aichele (Frankfurt) for helpful discussions.
\end{acknowledgments}


\begin{thebibliography}{10}
\bibitem{foot_blobprefactor}
 The (model depending) prefactors may be obtained from the intercept of the
  dilute and compact power-law slopes in Fig.~\ref{fig1}. Since $\Rend$ becomes
  compact more rapidly as $\Rgyr$, the blob size defined using $\Rend$ is
  slightly smaller than the one defined using $\Rgyr$. Note that due to the
  (generalized) Porod scattering at the chain perimeter (Fig.~\ref{fig4}), it
  is incorrect to determine the blob size $\xi$ by an Ornstein-Zernike fit to
  the intramolecular form factor $F(q)$ \cite{mai00}.
\bibitem{foot_othertheta}
 Similar scaling plots are obtained using different monomer pairs $n$ and
  $m=n+s-1$ as long as chain end effects giving rise to different contact
  exponents $\theta_i$ are avoided \cite{Duplantier89,MKA09,MWK10}.
\bibitem{foot_IntraInter}
 This relation between the intrachain size distribution and the interchain pair
  correlation function is also of relevance, e.g., for semidilute polymer
  solutions in $d=3$ dimensions as may be seen from Ref.~\cite{MM00}. A
  difference is here, however, that for $d=3$ only the dilute limit and
  distances $r \ll \xi$ probing the structure within the blob are characterized
  by a critical exponent $\thetatwo$, while for $r \gg \xi$ the intrachain and
  related interchain correlations are perturbative ($\thetatwo=0$), i.e.
  correspond to analytic deviations with respect to the Gaussian chain model
  \cite{WCX11}.
\bibitem{foot_acutoff}
 The specific value $a=1.56$ is motivated by the first minimum of the pair
  distribution function which varies only weakly with $\rho$. For not too high
  densities the scaling of $\ninter$ is found to be robust under variation of
  $a$. The quality of the $\rho$-scaling (but not of the scaling with respect
  to $N$) depends somewhat on $a$ for $\rho \ge 0.75$ where additional physics
  due to the discrete beads becomes relevant.
\bibitem{foot_blobsurface}
 The fraction of blobs on the perimeter of the compact chains can be computed
  directly by partitioning the chains in curvilinear subchains of length $g
  \approx 1/\rho^2$ and counting the number of subchains with at least one
  monomer in contact with a monomer from another chain. This fraction is found
  to decay as $(N/g)^{-3/8}$ (not shown).
\end{thebibliography}

\providecommand{\refin}[1]{\\ \textbf{Referenced in:} #1}

\clearpage
\newpage
\begin{figure}[t]
\centerline{\resizebox{0.9\columnwidth}{!}{\includegraphics*{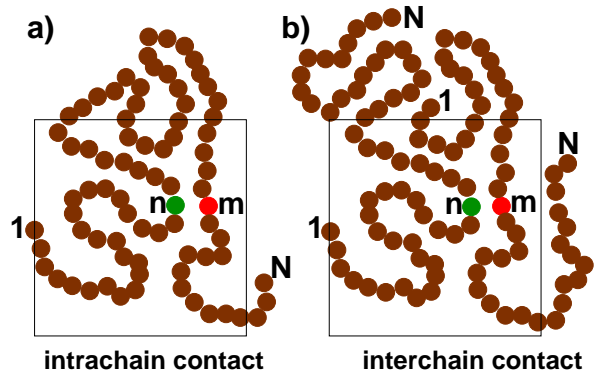}}}
\caption{TOC figure:
Key scaling argument relating the intrachain size distribution $G(r,s\approx N)$
to the interchain pair distribution function $\ginter(r)$. It states that
it is irrelevant of whether the neighborhood around the
reference monomer $n$ is penetrated by a long loop of the same chain as in panel (a)
or by another chain with a center of mass at a typical distance $R$ as shown in panel (b).
\label{figTOC}
}
\end{figure}

\end{document}